\documentclass[superscriptaddress,floatfix,showpacs,10pt,twocolumn]{revtex4}
\usepackage{amsmath,amsfonts,amssymb,graphicx,epsfig,color,times,bbm}

\begin{document}

\title{Quantum computation with quantum-dot spin qubits inside a cavity}

\author{Ping Dong}
\email{dongping9979@163.com}

\author{Ming Yang}

\affiliation{Key Laboratory of Opto-electronic Information
Acquisition and Manipulation, Ministry of Education, School of
Physics {\&} Material Science, Anhui University, Hefei, 230039, P R
China}

\author{Zhuo-Liang Cao}
\email{zhuoliangcao@gmail.com}

\affiliation{Key Laboratory of Opto-electronic Information
Acquisition and Manipulation, Ministry of Education, School of
Physics {\&} Material Science, Anhui University, Hefei, 230039, P R
China}

\affiliation{The School of Science, Hangzhou Dianzi University,
Hangzhou, 310018, P R China}

\begin{abstract}
Universal set of quantum gates are realized from the conduction-band
electron spin qubits of quantum dots embedded in a microcavity via
two-channel Raman interaction. All of the gate operations are
independent of the cavity mode  states, \emph{i.e.}, insensitive to
the thermal cavity field. Individual addressing and effective switch
of the cavity mediated interaction are directly possible here.
Meanwhile, gate operations also can be carried out in parallel. The
simple realization of needed interaction for selective qubits makes
current scenario more suitable for scalable quantum computation.
\end{abstract}

\pacs{03.67.Lx, 03.65.Ud, 68.65.Hb, 42.55.Sa}

\maketitle

Quantum computer can provide a possible alternative for resolving
certain hard problems in comparison with classical computer with the
help of the principle of coherent superposition and quantum
entanglement \cite{nielsen}. Solid state system has been generally
accepted to be the most promising hardware for quantum computation
since it can be easily integrated into large quantum networks. With
the development of fabrication and manipulation technologies in
semiconductor quantum dots, quantum computation based on this system
has attracted much attention. In a quantum dot system, decoherence
is still an important and challenging issue. However localized
electron spin state has relatively long decoherence time, so it is
more suitable as qubit. The realization of gate operations on
arbitrary two qubits is another challenge in solid state system. In
order to conquer this problem, Imamoglu and coworkers introduced the
quantum dot cavity QED scheme \cite{1} where the cavity mode can be
used as a data bus for long-distance information transfer and fast
coupling of arbitrary two qubits. In addition, this setup can
support parallel quantum logic gate operations. From then on, many
schemes adopt quantum dots embedded in cavity have been presented
\cite{2,3,4,5}.

In this paper, we propose a scenario for realizing quantum
computation via a two-channel Raman interaction of quantum dots
embedded in a  microcavity. Qubits are encoded on the
conduction-band spin states of semiconductor quantum dot. The
valence-band state is used as an auxiliary state, which can be
adiabatically eliminated. The decoherence time of qubits is long
enough to complete indispensable gate operations. The two-channel
Raman interaction model has been generally acceptable as an better
alternative to the single-channel one in atomic cavity QED system
\cite{7,8,9,10,6} as the easy realization of needed interactions.
Therefore, it is very significative to generalized the two-channel
Raman interaction model to quantum dot cavity QED system for solid
quantum computation. In fact, in comparison with atomic cavity QED,
quantum-dot cavity QED is more superior because quantum dots are
always fixed in a cavity, thus the scale up of the solid nature
system is quite straightforwardly. Meanwhile, individual addressing
of quantum dot qubits, which is of great importance for scalable
quantum computation, is directly possible taken into account the
fact that quantum dot is generally fabricated as a mesoscopic
quantum system.

\begin{figure}
\includegraphics[scale=1.01,angle=0]{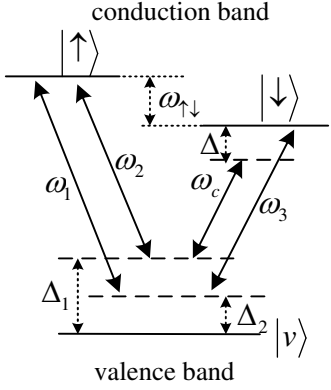}
\caption{The relevant energy levels of a single quantum dot .
$|\uparrow\rangle$ and $|\downarrow\rangle$ denote the spin up and
down states of the conduction-band electron, respectively, and
$|v\rangle$ denotes the valence-band state. $\omega_{j}$ $(j=1,2,3)$
are the frequencies of classical laser fields and $\omega_{c}$ is
the frequency of the cavity field.}\label{fig1}
\end{figure}

We consider $N$ III-V semiconductor quantum dots embedded in a
microcavity. All of the quantum dots are doped such that each
quantum dot has a single conduction-band electron and a full valence
band. Under the condition of quantum confinement, the
conduction-band electron is always in the ground state orbital. The
qubit is encoded on the conduction-band state $|\uparrow\rangle$ and
$|\downarrow\rangle$ by a uniform magnetic field. The relevant
energy levels of every quantum dot can be simulated as a three-level
configuration as shown in Fig. (\ref{fig1}).
$\hbar\omega_{\uparrow}$, $\hbar\omega_{\downarrow}$ and
$\hbar\omega_{v}$ are energies of the state $|\uparrow\rangle$,
$|\downarrow\rangle$ and $|v\rangle$, respectively, and
$\omega_{\uparrow\downarrow}=\omega_{\uparrow}-\omega_{\downarrow}$.
$\omega_{j}$ with $j=1,2,3$ are the frequencies of classical laser
fields and $\omega_{c}$ is the frequency of the cavity field.
$\Delta_{1}$, $\Delta_{2}$ and $\Delta$ are three detunings.
Assuming that $\Delta_{1}=\omega_{\uparrow}-\omega_{v}-\omega_{2}
=\omega_{\downarrow}-\omega_{v}-\omega_{c}-\Delta$ and
$\Delta_{2}=\omega_{\uparrow}-\omega_{v}-\omega_{1}
=\omega_{\downarrow}-\omega_{v}-\omega_{3}$, so we have
$\omega_{\uparrow\downarrow}+\Delta=\omega_{2}-\omega_{c}$ and
$\omega_{\uparrow\downarrow}=\omega_{1}-\omega_{3}$.  Every quantum
dot is off-resonant excited via two Raman channels by using
classical laser fields and the microcavity. One channel consists of
laser fields 1 and 3, the other consists of laser field 2 and the
microcavity field.  The total system contains $N$ quantum dots, a
microcavity and $3N$ classical laser fields, the Hamiltonian of
which can be described as (assuming $\hbar=1$)
\begin{subequations}
\label{1}
\begin{equation}
H=H_{0}+H_{int},
\end{equation}
\begin{equation}\label{1b}
H_{0}=\sum_{i=1}^{N}(\omega_{\uparrow}\sigma^{i}_{\uparrow\uparrow}+
\omega_{\downarrow}\sigma^{i}_{\downarrow\downarrow}+\omega_{v}\sigma^{i}_{vv})+
\omega_{c}a^{\dag}a,
\end{equation}
\begin{eqnarray}\label{int}
H_{int}&=&\sum_{i=1}^{N}\bigg[(\Omega_{1} e^{-i\omega_{1}t} +
\Omega_{2} e^{-i\omega_{2}t})\sigma^{i}_{\uparrow v} \nonumber\\
&\quad& +(\Omega_{3} e^{-i\omega_{3}t} + ga)\sigma^{i}_{\downarrow
v} +H.c.\bigg],
\end{eqnarray}
\end{subequations}
where $\Omega_{j}$ with $j=1,2,3$ are Rabi frequencies of classical
fields, $g$ is coupling constant of the microcavity mode and each
quantum dot with index number $i$, and $\sigma_{mn}=|m\rangle\langle
n|$ $(m,n=\uparrow,\downarrow,v)$.  In writing Eq. (\ref{int}), we
have assumed that $\Omega^{i}_{j}=\Omega_{j}$ and $g^{i}=g$.

The interaction Hamiltonian (\ref{int}) can be rewritten, in the
interaction picture with respect to  (\ref{1b}), as
\begin{eqnarray}
H_{I}&=&\sum_{i=1}^{N}\bigg[\Omega_{2}\sigma^{i}_{\uparrow
v}e^{i\Delta^{i}_{1}t}+(\Omega_{1}\sigma^{i}_{\uparrow
v}+\Omega_{3}\sigma^{i}_{\downarrow
v})e^{i\Delta^{i}_{2}t}\nonumber\\&+&  g a\sigma^{i}_{\downarrow
v}e^{i(\Delta^{i}_{1}+\Delta^{i})t}+H.c.\bigg].
\end{eqnarray}

In the case of $\Delta_{1},\Delta_{2}\gg \Omega_{j}, g$ and
$\Delta_{1}-\Delta_{2}\gg \{\Delta,
\frac{(\Delta_{1}+\Delta_{2})\Omega_{1}\Omega_{2}}{2\Delta_{1}\Delta_{2}},
\frac{(\Delta_{1}+\Delta_{2})\Omega_{2}\Omega_{3}}{2\Delta_{1}\Delta_{2}},
\frac{(2\Delta_{1}+\Delta)\Omega_{1}g}{2\Delta_{1}(\Delta_{1}+\Delta)},
\frac{(2\Delta_{1}+\Delta)\Omega_{3}g}{2\Delta_{1}(\Delta_{1}+\Delta)}\}$,
the valence-band state can be adiabatically eliminated \cite{8}. We
can then obtain an effective Hamiltonian by using rotating-wave
approximation
\begin{eqnarray}
\label{3}
H_{e}^{(1)}&=&\sum_{i=1}^{N}\bigg[\frac{\Omega_{1}\Omega_{3}}{\Delta^{i}_{2}}\left(\sigma^{i}_{\uparrow\downarrow
}+\sigma^{i}_{\downarrow\uparrow }\right)+\frac{
g\Omega_{2}}{2}\left(\frac{1}{\Delta^{i}_{1}}+\frac{1}{\Delta^{i}_{1}+\Delta^{i}}\right)\nonumber\\&&
\left(a^{\dag}\sigma^{i}_{\uparrow\downarrow}e^{-i\Delta^{i}
t}+a\sigma^{i}_{\downarrow\uparrow}e^{i\Delta^{i} t}\right)\bigg],
\end{eqnarray}
where we  have neglected the ac-Stark energy shift, which can be
easily compensated \cite{shift} by an addition laser field
dispersively coupled to an energy level outside the qubit space in
real experimental implementation.

For simplification of calculation, we choose a new computational
basis
$|\pm\rangle^{i}=\frac{1}{\sqrt{2}}(|\uparrow\rangle^{i}\pm|\downarrow\rangle^{i})$.
We can rewrite the effective Hamiltonian (\ref{3}) as
\begin{eqnarray}
\label{4}
H_{e}^{(2)}&=&\sum_{i=1}^{N}\bigg[A\bigg(\frac{2S^{i}_{z}-S^{i}_{-}+S^{i}_{+}}{4}a^{\dag}e^{-i\Delta^{i}
t}\nonumber\\&\quad&+\frac{2S^{i}_{z}+S^{i}_{-}-S^{i}_{+}}{4}ae^{i\Delta^{i}
t}\bigg)+BS^{i}_{z}\bigg],
\end{eqnarray}
where $A=\frac{
g\Omega_{2}}{2}\left(\frac{1}{\Delta^{i}_{1}}+\frac{1}{\Delta^{i}_{1}+\Delta^{i}}\right)$,
$B=\frac{2\Omega_{1}\Omega_{3}}{\Delta^{i}_{2}}$,
$S_{+}=|+\rangle\langle-|$, $S_{-}=|-\rangle\langle+|$ and
$S_{z}=\frac{1}{2}(|+\rangle\langle+|-|-\rangle\langle-|)$.

Assume that $B\gg \Delta^{i}, A$ and in the $S_{z}$ framework
$H'_{0}=BS^{i}_{z}$, the Hamiltonian (\ref{4}) can be reduced to
\begin{eqnarray}
H_{e}&=&\sum_{i=1}^{N}\left[\frac{A}{2}\left(a^{\dag}e^{-i\Delta^{i}
t}+ae^{i\Delta^{i} t}\right)S^{i}_{z}\right]\nonumber\\
&=& \sum_{i=1}^{N}\left[\frac{A}{2}\left(a^{\dag}e^{-i\Delta^{i} t}+
ae^{i\Delta^{i} t}\right)\left(\sigma^{i}_{\uparrow\downarrow
}+\sigma^{i}_{\downarrow\uparrow }\right)\right].
\end{eqnarray}

For the implementation of quantum computation, the most important
steps should be the realization of  a set of universal quantum
logical gates, \emph{i.e}., two-qubit logic gate, controlled-not
gate or controlled phase shift, and arbitrary single-qubit
rotations. Here we first introduce the scenario for implementing a
controlled phase shift.  We turn on three classical laser fields
$\omega_{j}$ on quantum dots $m$ and $n$, let quantum dot $m$
interacts with $n$ via the virtue excited cavity mode under the
condition of $\Delta^{m}=\Delta^{n}=\Delta$. The time evolution
operator for this system can be expressed as this form
\begin{equation}
\label{6}
U=e^{-i\alpha(t)(\sum_{l}S^{l}_{z})^{2}}e^{-i\beta(t)\sum_{l}S^{l}_{z}a}
e^{-i\gamma(t)\sum_{l}S^{l}_{z}a^{\dag}},
\end{equation}
where $l=m,n$. The coefficients $\alpha(t)$, $\beta(t)$ and
$\gamma(t)$ can be calculated by Schr\"{o}dinger equation as
\cite{11,12}
\begin{subequations}
\begin{equation}
\beta(t)=\int^{t}_{0}\frac{A}{2}e^{i\Delta
t'}dt'=\frac{A}{2i\Delta}(e^{i\Delta t}-1),
\end{equation}
\begin{equation}
\gamma(t)=\int^{t}_{0}\frac{A}{2}e^{-i\Delta
t'}dt'=\frac{-A}{2i\Delta}(e^{-i\Delta t}-1),
\end{equation}
\begin{eqnarray}
\alpha(t)&=&i \int^{t}_{0}\beta(t')\frac{A}{2}e^{-i\Delta
t'}dt'\nonumber\\&=&\frac{A^{2}}{4\Delta}\left[t-\frac{i}{\Delta}(e^{i\Delta
t}-1)\right].
\end{eqnarray}
\end{subequations}

Setting $\Delta t=2\pi$ results in $ \beta(t)=\gamma(t)=0$ and
$\alpha(t)=\frac{A^{2}}{4\Delta\hbar^{2}}t$, and thus the total
evolution operator of the system becomes
\begin{equation}
U_{m,n}=e^{-iBt(\sum_{l}S^{l}_{z})}e^{-i\frac{A^{2}}{4\Delta}t(\sum_{l}S^{l}_{z})^{2}},
\end{equation}
so that the state evolutions of $|++\rangle_{mn}$,
$|+-\rangle_{mn}$, $|-+\rangle_{mn}$ and $|--\rangle_{mn}$ are
\begin{subequations}
\label{9}
\begin{equation}
|++\rangle_{mn}\rightarrow
e^{-i(B+\frac{A^{2}}{4\Delta})t}|++\rangle_{mn},
\end{equation}
\begin{equation}
|--\rangle_{mn}\rightarrow
e^{-i(\frac{A^{2}}{4\Delta}-B)t}|--\rangle_{mn},
\end{equation}
\begin{equation}
|+-\rangle_{mn}\rightarrow |+-\rangle_{mn},
\end{equation}
\begin{equation}
|-+\rangle_{mn}\rightarrow |-+\rangle_{mn}.
\end{equation}
\end{subequations}
If we parameterize the  interaction time and Rabi frequencies as
\begin{subequations}
\begin{equation}
Bt= 2k\pi+\frac{\pi}{2},
\end{equation}
\begin{equation}
\frac{A^{2}}{4\Delta}t=\frac{\pi}{2},
\end{equation}
\end{subequations}
where $k=0,\pm1,\pm2,\cdots$, which correspond that all parameters
should satisfy the conditions
\begin{subequations}
\label{11}
\begin{equation}
\Delta=\frac{g\Omega_{2}}{2}\left(\frac{1}{\Delta_{1}}+\frac{1}{\Delta_{1}+\Delta}\right),
\end{equation}
\begin{equation}
\frac{4\pi\Omega_{1}\Omega_{3}}{\Delta\Delta_{2}}=2k\pi+\frac{\pi}{2},
\end{equation}
\end{subequations}
then, the evolution of states of Eq. (\ref{9}) becomes
\begin{subequations}
\begin{equation}
|++\rangle_{mn}\rightarrow -|++\rangle_{mn},
\end{equation}
\begin{equation}
|--\rangle_{mn}\rightarrow |--\rangle_{mn},
\end{equation}
\begin{equation}
|+-\rangle_{mn}\rightarrow |+-\rangle_{mn},
\end{equation}
\begin{equation}
|-+\rangle_{mn}\rightarrow |-+\rangle_{mn}.
\end{equation}
\end{subequations}
Obviously, this is a standard controlled phase shift transformation
under the basis $\{|+\rangle, |-\rangle\}$. If we rotate the basis
with an angle $\theta=\pi/4$, the controlled phase gate is then
implemented in the qubit space.

Similarly, we can realized the controlled phase shift between
arbitrary two spin qubits of the $N$ quantum dots. The cavity
mediated interaction can be implemented on selective qubits, this
can be achieved by turning on/off the external driving lasers on
certain qubits. In fact, the qubits interaction is mediated by
virtual exchange of photons with the cavity, which requires the
qubits are "degenerate" with each other \cite{switch}. Here,
"degenerate" should means the same effective detuning $\Delta$ for
considered qubits. When the qubits are non-degenerate, similar to
the argument in  \cite{switch}, the cavity mediated process can be
effectively turned off as it does not conserve energy. Therefore, we
also can carry out the controlled phase gate in parallel if we turn
on the two-channel Raman resonant on different pair of qubits
simultaneously with different pairs working in different detunings.
Then, cross-talk of different pairs can be effective neglected
provided that the difference of the working detunings is
considerately large, thus results in parallel computation a natural
merit in present scenario of quantum computation.

The cavity-state-free evolution (\ref{9}) is achieved by periodical
evolution ($\Delta T=2\pi$) of a near-resonant driving with detuning
$\delta$ and periodicity $T = 2\pi/\delta$ (in the rotating frame).
Physically, after periodical evolution following the path
$\mathcal{L}=\frac{A}{2\Delta}\left(1-e^{\text{i}\Delta t}\right)$,
the cavity state returns to its original phase space coordinates
with an additional geometric phase $\alpha(t)$ equivalent to the
area enclosed by the trajectory \cite{12}. So, the present gate
operation is also of geometric nature, which is generally believed
to be more robust against random operation errors.

Then we briefly introduce the single-qubit operations in this
system. As single-qubit operations is much faster than that of the
two-qubit case, we can simply consult the Raman process with laser
fields 1 and 3. Laser field 2 is turn off now, thus the cavity field
can be effectively eliminated. The net effect of the cavity field is
an additional ac Stark shift, inversely proportional to $\Delta_1$.
Considering the fact that $\Delta_1$ is the detuning of optical
frequencies, this term can be safely neglected. In this case, the
process only controlled by the single-channel Raman interaction,
consists of laser fields 1 and 3, and the effective Hamiltonian is
\begin{equation}
H'=\frac{\Omega_{1}\Omega_{3}}{\Delta_{2}}\left(\sigma_{\uparrow\downarrow
}+\sigma_{\downarrow\uparrow }\right),
\end{equation}
from which we can realize arbitrary single-qubit rotations by
choosing appropriate interaction time. The process is equivalent to
the single-channel Raman process with two classical laser fields of
Ref. \cite{1}. Single-qubit operations also can be implemented by
using external magnetic fields with different directions \cite{13}.

Next we discuss the feasibility of the current scenario. From Eq.
(\ref{11}), we can obtain $g=\frac{16}{4k+1}$ meV where $k=5$ can be
deserved in terms of the current experimental parameter $g\sim 0.5$
meV \cite{1}. We also can obtain $A=0.1g$, $B=0.4$ meV and
$\Delta=0.1g$. Choosing the typical parameters to satisfy
$\Omega_{j}=1$ meV, $\Delta_{2}=5$ meV \cite{5} and $\Delta_{1}=10$
meV, we will obtain $\Delta_{1}-\Delta_{2}=5$ meV, which satisfies
the first approximate condition $\Delta_{1}-\Delta_{2}\gg \{\Delta,
\frac{(\Delta_{1}+\Delta_{2})\Omega_{1}\Omega_{2}}{2\Delta_{1}\Delta_{2}},
\frac{(\Delta_{1}+\Delta_{2})\Omega_{2}\Omega_{3}}{2\Delta_{1}\Delta_{2}},
\frac{(2\Delta_{1}+\Delta)\Omega_{1}g}{2\Delta_{1}(\Delta_{1}+\Delta)},
\frac{(2\Delta_{1}+\Delta)\Omega_{3}g}{2\Delta_{1}(\Delta_{1}+\Delta)}\}$.
The second approximate condition $B\gg \Delta^{i}, A$ can be
satisfied automatically because $B=0.4$ meV is larger than
$A=\Delta= 0.1 g \sim 0.05$ meV.

The time required to complete a single-qubit rotation is about
$t_{s}\sim 10$ ps under above mentioned conditions, which is the
same as the single-channel Raman interaction process. The
implementation of two-qubit gate operation need about $t_{t}\sim
100$ ps, which is close to that in the case of sing-channel Raman
interaction process \cite{1}, where one needs twice single-channel
interactions on the two qubits and some single-qubit rotations for
realizing a controlled phase shift. However, in our scheme, we only
needs one two-channel Raman interaction on the two qubits without
the help of single-qubit rotations. Thus the two-channel Raman
interaction process is simpler for scalable quantum computation than
single-channel Raman interaction process in the quantum-dot spin
system.

Decoherence is a main obstacle in quantum information processing,
thus we should consider the relative magnitude of the decoherence
rates as compared to the gate-operation time. The coherent time of
conduction-band electrons is about 1 $\mu$s in doped quantum well
and bulk semiconductors \cite{1}. Recent experiment indicated that
the spin coherent time can reach 1.2 $\mu$s by using spin-echo
technology \cite{14}. Obviously the gate-operation time is much less
than spin coherent time. Generally, in cavity QED schemes one should
consider the cavity decay factor and thermal field, which may
introduce a decoherence mechanism. In our scheme, the cavity mode is
only virtually excited during the interaction, but the effective
decoherence time will still be on the order of 1 ns with the cavity
lifetime $\Gamma\sim10$ ps \cite{1}. Situation can still be better,
we can embed the quantum dots in a microdisk (or microchip)
structure \cite{1,5,switch,15} to enhance the couple of quantum dots
and a single photon. Meanwhile, we can improve the cavity quality
factor by using high-Q whispering gallery mode of a silica
microsphere \cite{17} as well as photonic-crystal microcavity mode
\cite{18}. Furthermore, the photon-number-dependent parts in the
evolution operator are canceled, thus our scheme is insensitive to
the thermal field. In addition, one can carry out the single- or
two-qubit operations in parallel, which can reduce the total
operation time comparing with the sequential individual gate
operation method.

Addressing and capture for particles are another issues for quantum
information processing. In current scheme, all of the quantum dots
are trapped in a microcavity, the positions of quantum dots are
fixed, so one does not need capture the quantum dots. We only need
consider the selective addressing problem, which has been
successfully demonstrated in experiments. In the case that the
number of quantum dots is small, every quantum dot can be addressed
selectively by a laser field from a fiber tip (near-filed
technology) \cite{1}. In the case of scalable quantum computation,
we should consult to the switch on/off technology, which has already
been considered above.

In conclusion, we have presented a scenario for realizing quantum
computation with quantum dots spins and microcavity by using a
two-channel Raman resonant interaction.  The two-channel Raman
resonant interaction model is more convenient than the previous
single-channel Raman process \cite{1}. The gate operations do not
depend on the state of cavity mode, \emph{i.e}., insensitive to the
thermal field, and the acquired phase is of geometric nature. The
effective switch method presented makes the selective qubits
interaction and parallel computation are both possible, which is
very important for scalable quantum computation. Detail discussions
show that the present set-up is also in the reach of current
technology.

\acknowledgments

P. Dong thanks Z.-Y. Xue for many useful discussions. This work is
supported by National Natural Science Foundation of China (NSFC)
under Grant Nos: 60678022 and 10704001, the Specialized Research
Fund for the Doctoral Program of Higher Education under Grant No.
20060357008, Anhui Provincial Natural Science Foundation under Grant
No. 070412060, the Talent Foundation of Anhui University, Anhui Key
Laboratory of Information Materials and Devices (Anhui University),
and the Doctorial Innovation Research Plan Fund from Anhui
University under Grant No. 20072007.

\end{document}